\documentclass[conference]{IEEEtran}
\usepackage{todonotes}
\usepackage{amsmath,amssymb,amsfonts}
\usepackage{algorithmic}
\usepackage{graphicx}
\usepackage{textcomp}
\usepackage{xcolor}
\usepackage{braket}
\usepackage{flushend}
\usepackage{caption}
\usepackage{subcaption}
\usepackage[utf8]{inputenc}
\usepackage[T1]{fontenc}
\usepackage{bm}
\usepackage{float}
\usepackage{layouts}
\usepackage{hyperref}
\usepackage{booktabs}
\usepackage{listings}
\usepackage{color}
\lstdefinestyle{mystyle}{
    backgroundcolor=\color{backcolour},   
    commentstyle=\color{codegreen},
    keywordstyle=\color{magenta},
    numberstyle=\tiny\color{codegray},
    stringstyle=\color{codepurple},
    breakatwhitespace=false,         
    breaklines=true,                 
    captionpos=b,                    
    keepspaces=true,                 
    numbers=left,                    
    numbersep=5pt,                  
    showspaces=false,                
    showstringspaces=false,
    showtabs=false,                  
    tabsize=2
}
\usepackage[final]{pdfpages}
\newcommand\blfootnote[1]{%
  \begingroup
  \renewcommand\thefootnote{}\footnote{#1}%
  \addtocounter{footnote}{-1}%
  \endgroup
}

\def\BibTeX{{\rm B\kern-.05em{\sc i\kern-.025em b}\kern-.08em
   T\kern-.1667em\lower.7ex\hbox{E}\kern-.125emX}}
\usepackage{cite}

\DeclareCaptionLabelSeparator{periodspace}{.\quad}
\usepackage{pgf}
    
\begin{document}
\bstctlcite{IEEEexample:BSTcontrol}


\title{Phase-Based Approaches for Rapid Construction of Magnetic Fields in NV Magnetometry}

\author{\IEEEauthorblockN{Prabhat Anand\IEEEauthorrefmark{1}, Ankit Khandelwal\IEEEauthorrefmark{1}, Achanna Anil Kumar\IEEEauthorrefmark{1}, M Girish Chandra\IEEEauthorrefmark{1}, Pavan K Reddy\IEEEauthorrefmark{1},\\ Anuj Bathla\IEEEauthorrefmark{3}, Dasika Shishir \IEEEauthorrefmark{2}, Kasturi Saha\IEEEauthorrefmark{2}\IEEEauthorrefmark{4}\IEEEauthorrefmark{5}}
\IEEEauthorblockA{\IEEEauthorrefmark{1}\textit{TCS Research, Tata Consultancy Services Ltd, Bengaluru}, India}
\IEEEauthorblockA{\IEEEauthorrefmark{2}\textit{Department of Electrical Engineering, Indian Institute of Technology Bombay (IITB), Mumbai}, India}
\IEEEauthorblockA{\IEEEauthorrefmark{3}\textit{Centre for Research in Nanotechnology and Science (CRNTS), Department of Electrical Engineering, IITB}, India}
\IEEEauthorblockA{\IEEEauthorrefmark{4}\textit{Centre of Excellence in Quantum Information, Computing Science and Technology, IITB}, India}
\IEEEauthorblockA{\IEEEauthorrefmark{5}\textit{Centre of Excellence in Semiconductor Technologies (SemiX), IITB,} India }
\vspace{-10mm}
}

\maketitle
\begin{abstract}

With the second quantum revolution underway, quantum-enhanced sensors are moving from laboratory demonstrations to field deployments, providing enhanced and even new capabilities. Signal processing and operational software are becoming integral parts of these emerging sensing systems to reap the benefits of this progress. This paper looks into widefield Nitrogen Vacancy Center-based magnetometry and focuses on estimating the magnetic field from the Optically Detected Magnetic Resonances (ODMR) signal, a crucial output for various applications. Mapping the shifts of ODMR signals to phase estimation, a computationally efficient approaches are proposed. Involving Fourier Transform and Filtering as pre-processing steps, the suggested approaches involve linear curve fit or complex frequency estimation based on well known super-resolution technique Estimation of Signal Parameters via Rotational Invariant Techniques (ESPRIT). The existing methods in the quantum sensing literature take different routes based on Lorentzian fitting for determining magnetic field maps. To showcase the functionality and effectiveness of the suggested techniques, relevant results, based on experimental data are provided, which shows a significant reduction in computational time with the proposed method over existing methods.
\end{abstract}


\section{Introduction}
\blfootnote{Correspondence should be addressed to anand.prabhat@tcs.com}
It is widely believed that continuous advances in modern science and technology rely heavily on precise sensing and atomic level magnetism control \cite{Lee19}. A very promising quantum sensing approach, signifying the progression from fundamental research to practical application, is based on the negatively charged nitrogen vacancy (NV) centers in the diamond crystal \cite{sum1,sum4,sum8}. They provide high magnetic sensitivity in ambient conditions with large dynamic range and high spatial resolution \cite{sum1}. This technology can be applied to efficiently measure temperature, electric and magnetic fields, and pressure. The magnetic field detection of the NV centers relies on the magnetic interaction of the electron spin via the Zeeman effect \cite{sum8, sum2}. The NV spin state can be optically initialized, suitably governed using Microwave (MW) fields, and read out optically with extended coherence periods \cite{sum1}. 

\begin{figure}[ht]
    \centering
    \includegraphics[width=\linewidth]{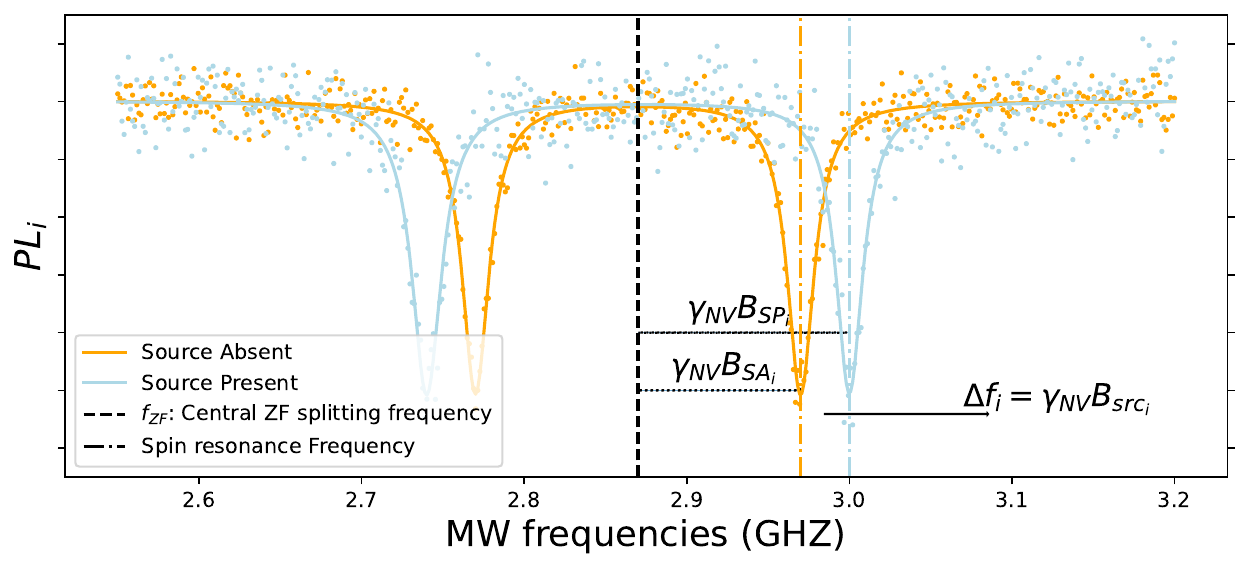}
    \caption{A typical ODMR has PL recorded over a band of MW frequencies. 
    Here, two ODMRs for a single $i^{th}$ NV axis are shown in the presence and absence of a magnetic source demonstrating magnetic field-dependent shifts in resonant frequencies; elaborated in (\ref{bias}). However, since a diamond crystal allows four different orientations of NV centers, eight dips are expected in the ODMR of NV ensemble system, enabling the construction of a 3D vectorial magnetic field through wide-field microscopy using full ODMR. 
    }
    \label{odmrfig}
\end{figure}

The advancement of magnetic sensor technology that can simultaneously achieve both high magnetic sensitivity and high spatial resolution with large dynamic range is becoming increasingly vital across a variety of fields, from solid-state physics to life sciences \cite{sum1,sum3, Lee19}. For the work presented here, we consider a widefield magnetometer consisting of an ensemble of NV centers, that is, a dense layer of NV centers induced near the surface of diamond \cite{setup}. A typical widefield experimental setup can be seen in \cite[Fig.~2.11]{marwa} and  in \cite[Fig.~1]{setup}. One of the important steps for many applications involving NV magnetometry is a good construction of 2D maps of 3D vectorial magnetic fields \cite{sum7}. Each pixel value of these maps is typically determined from the ODMR “signal” (Fig. \ref{odmrfig}) obtained from the experimental setup.  

\begin{figure*}[ht]
\centering
\begin{minipage}{0.3\linewidth}
   \includegraphics[width=1\linewidth]{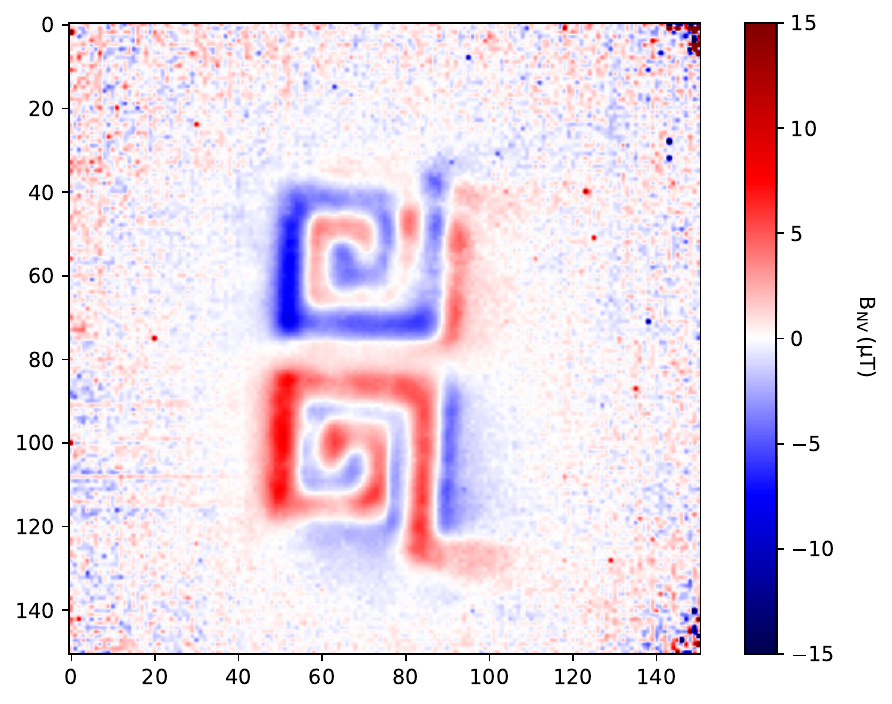}
   \subcaption{Existing Non-linear Curve fitting}
   \label{noncur}
\end{minipage}
\begin{minipage}{0.3\linewidth}
   \includegraphics[width=1\linewidth]{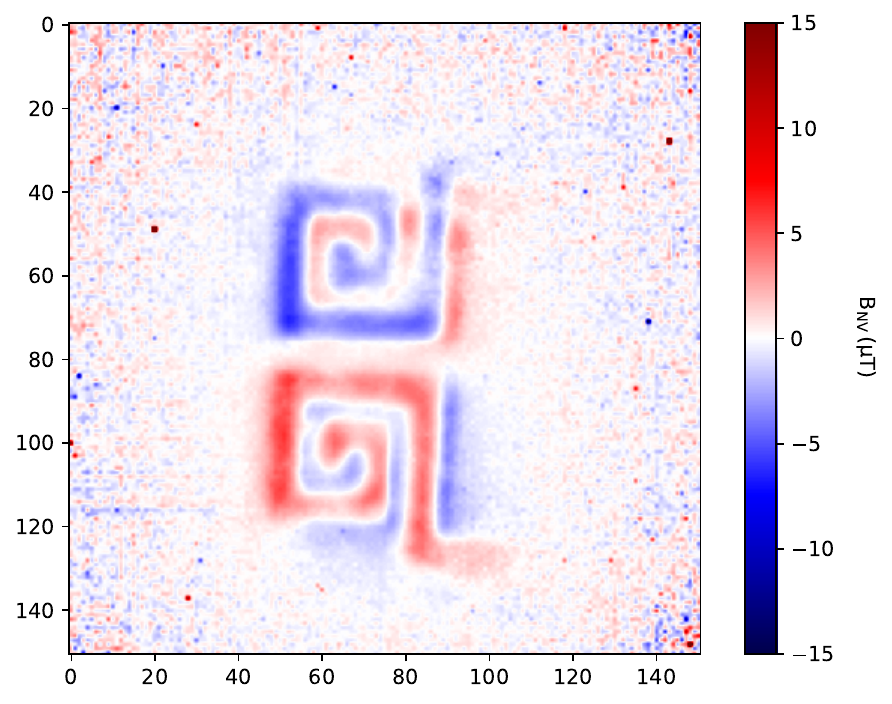}
   \subcaption{Proposed Linear Curve fitting approach}
   \label{lincur}
\end{minipage}
\begin{minipage}{0.3\linewidth}
   \includegraphics[width=1\linewidth]{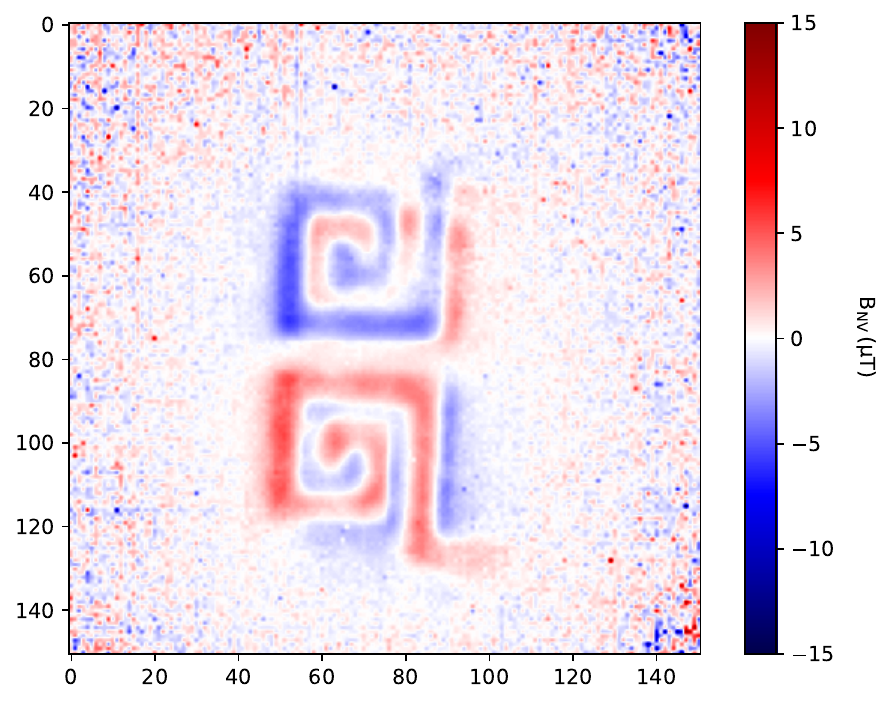}
   \subcaption{Proposed ESPRIT approach}
   \label{esprit}
\end{minipage}
\caption{Construction of one component of Magnetic Field from ODMR: (\subref{noncur}) The TRF algorithm implemented in scipy is used to fit a Lorentzian. (\subref{lincur}) Phase determination incorporating linear curve fit. (\subref{esprit}) Phase determination incorporating ESPRIT.}
\label{map}
\end{figure*}

The characteristic ODMR in widefield imaging mode is produced when red photoluminescence (PL) of NV-ensembles is recorded by a conventional scientific camera (say, Charge Coupled Device (CCD) camera) in the presence of a MW field with varying frequencies \cite{setup}. Each frequency step results in a different 2D frame of PL. The values in the consecutive frames for each pixel, put together, result in the ODMR signal, which is basically a signal in the frequency domain (Fig. \ref{odmrfig}). A popular way of measuring the magnetic field of a source is by tracking the shifts in the resonant frequency of the ODMR signals in the presence and absence of the source \cite{setup, sum3}. With appropriate processing, a precise imaging of the 3D magnetic field is possible.

There are different propositions \cite{marwa,sum3}, in the existing literature to process ODMR signal to construct magnetic maps. It is important to note at this juncture that the measured ODMR signal suffers from noise corruption and other distortions depending on the setup \cite{noise1, noise2, noise3, noise4, noise5}. There is also an effect due to the separation between the magnetic source and the NV-induced diamond slab plane, known as stand-off distance. Viewing from a signal processing perspective one can think of an “ideal” ODMR signal passing through a system that introduces artifacts which is further affected by noise. This opens up a relatively new territory for the signal-processing community to contribute to the exciting area of quantum-enhanced sensing. 

The paper is organized as follows: in Section \ref{secmet}, we propose a novel technique by adopting methods of signal processing and suitably modifying them to arrive at ways to construct the magnetic field from the ODMR signal. Such an approach does not exist in the context of quantum sensing literature to the best of our knowledge. The relevant results captured in Section \ref{secres} demonstrate the applicability and usefulness of the technique including $\sim 70\times$ improvement in processing speed compared to the existing method on real data. It is followed by concluding remarks with outlook for Signal processing community in Section \ref{seccon}. 
\section{Magnetic Field Construction from ODMR}\label{secmet}
As the physics and technological advancements of NV-center based magnetometry have been extensively documented in other works \cite{marwa, sum1, sum5, sum6}, this paper will not address those topics. Restricting to the ODMR signal, Fig. \ref{odmrfig} depicts data points of a section of a typical ODMR signal.
In absence of any magnetic field, NV centers have a degenerate excited state having an energy gap, corresponding to roughly $f_{ZF}=D_{gs}=2.87$ GHz, called zero-field (ZF) splitting frequency (see Fig. \ref{odmrfig}) giving a single dip in ODMR centered around it.
The number of dips increases in the presence of magnetic field with every two dips centered around one of the four magnetic field component ($B_{NV_i}$)-dependent resonant frequencies $f_i$ given in (\ref{gap}). 
Once this intensive task of determining the central frequencies $f_i$ is done from the ODMR, by tracking these changes in $f_i$, one can determine changes in the external fields. 
Hence, two sets of measurements are carried out for magnetic field construction (\ref{gap}), that is, an MW frequency sweep is carried out in the absence of any sample (SA) when $B_{NV_i} = B_{SA_i}$ and is subtracted from $B_{NV_i} = B_{SP_i}$ that is measured in the presence of the magnetic sample (SP) (\ref{bias}).
\begin{align}
    \Delta f_{i} &= | f_{{SP}_i} - f_{{SA}_i} | \notag\\ &= | \left( D_{gs}+\gamma_{NV} B_{SP_i} \right) - \left( D_{gs}+\gamma_{NV} B_{SA_i} \right) |\notag\\
    &= |\gamma_{NV} \left( B_{SP_i} - B_{SA_i}  \right) | = \gamma_{NV} B_{src_i}
    \label{gap} \\
    \text{given}\ \ \  B_{SP_i} &= B_{bias_i} + B_{src_i}, \ \notag \\ 
    B_{SA_i} &= B_{bias_i} \label{bias}
\end{align}

where $B_{bias_i}$ is bias magnetic field and $B_{src_i}(\ll B_{bias_i}) $ is magnetic field due to source along $i^{th}$ NV axis, $i\in\{1,2,\cdots ,8\}$ and $\gamma_{NV}\ (= 28\, kHz/\mu T)$ is the gyromagnetic ratio of NV center, a fundamental constant. The challenge however is to precisely determine the central frequency $f_i$ from ODMR signal of each pixel. 

\subsection{Few Remarks on Existing Methodologies}
\textit{Lorentzian-fitting Approach (LF)}:
To determine the central frequency of the dip in ODMR, typically Lorentzian (\ref{lorentz}) is fitted onto the full ODMR spectrum acquired, due to prevalence of Lorentzian broadening in Continuous-wave-ODMR \cite{lorentz}, using different multi-parameter optimization algorithms.  
\begin{align}
    f(\nu) = 1- \sum_{i=1}^8 C_i \left( \frac{\gamma^2}{4(\nu_{i}-\nu)^2+\gamma^2 }\right)
    \label{lorentz}
\end{align}

From this non-linear fitting, one can obtain the values of central frequencies which, in turn, give the value of external magnetic field using (\ref{bias}). A variation to this method, where the slope of Lorentzian is fitted, can be found in \cite{setup}. In the following section, we propose alternative efficient approaches based on signal processing methods.
\subsection{Proposed Phase-based Approaches}
To estimate the magnetic field $B_{src_i}$, from now onwards, we consider a portion of full ODMR around a single resonant frequency. Let $F_{SA}(f)$ and $F_{SP}(f)$ denote the ODMR with source absent and source present respectively (Fig. \ref{odmrfig}). Now, $F_{SP}(f)$ can be visualized, from a theoretical point of view, as \textit{frequency shifted signal} of $F_{SA}(f)$ by $B_{src_{i}}$. 
The actual experimental data brings some dissimilarity among the signals, but even then shifted property is a good first approximation.
Hence, we can express $F_{SP}(f) = F_{SA}(f -\Delta f_i)=  F_{SA}(f -\gamma_{NV}B_{src_{i}})$. Let the inverse Fourier Transform \cite{vetterli2014foundations} (IFT) of $F_{SA}(f)$, $\mathcal{F}^{-1}(F_{SA}(f)) = f_{SA}(t)$. Using the identities of IFT \cite{vetterli2014foundations},  $f_{SP}(t) = \mathcal{F}^{-1}(F_{SP}(f))$ can be expressed as
\begin{equation}
    f_{SP}(t) = f_{SA}(t)e^{j2\pi t \Delta f_i}.
    \label{eq:ift_fsp}
\end{equation}
The ODMR $F_{SA}(f)$ which is measured without the source can be obtained easily providing $f_{SA}(t)$; hence
\begin{equation}
    f_{B_{src}}(t) = \frac{f_{SP}(t)}{f_{SA}(t)} = e^{j2\pi t \Delta f_i}. 
    \label{eq:fbsrc}
\end{equation}
From (\ref{eq:fbsrc}) notice that the unknown $\Delta f_i = \gamma_{NV} B_{src_i}$ resides in the phase of $f_{B_{src}}(t)$.
Observing from Fig. \ref{odmrfig} and from typical NV-centers systems, $F_{SA}(f)$ is a slowly varying signal and hence $f_{SA}(t)$ is a lowpass signal in the time domain with most of the energy concentrated at the \textit{low-time} band region (analogous to standard low-frequency band signal).  
Hence, before further processing we filter the signal as  $f^{(filt)}_{B_{src}}(t) = f^{(filt)}(t)f_{B_{src}}(t)$, where $f^{(filt)}(t)$ is defined as

\begin{equation}
f^{(filt)}(t) =
    \begin{cases}
    1, & t \in [-T_t,+T_t] \\
    0, & \textrm{otherwise}
    \end{cases}
\end{equation}

We now use the following two approaches to estimate $B_{src_i}$ from $f^{(filt)}_{B_{src}}(t)$. 

\subsubsection{Linear Curve Fitting based Approach} We first unwrap $f^{(filt)}_{B_{src}}(t)$, which can be expressed as
\begin{equation}
  g(t) = -j\ln{(f^{(filt)}_{B_{src}}(t))} =  2\pi t \gamma_{NV}B_{src_i}.
   \label{eq:unwrap_signal}
\end{equation}
We then do linear curve fitting for the measurements $g(t)$ in the region $t \in [-T_t,+T_t]$ using the simple least squares approach. The slope of this line shall provide an estimate of $B_{src_i}$. 

\subsubsection{Super-resolution ESPRIT approach} 

Observe that estimating $\Delta f_i$ from $f^{(filt)}_{B_{src}}(t)$ is analogous to a complex frequency estimation (see (\ref{eq:fbsrc})). Hence, alternative to the above approach, we obtain  $B_{src_i}$ by employing complex frequency estimation techniques. In order to estimate with  a very good resolution, we employ the well known super-resolution techniques like ESPRIT \cite{moon2000mathematical} for frequency estimation. 

\subsection{Comparison between Lorentzian Fitting and Phase-based Approaches}

Notice from \eqref{eq:fbsrc}, the existing LF-based approach requires a computationally intensive curve-fitting of a non-linear multi-variate function. Bypassing this, in the proposed approach requires Fourier Transform (FT) computation of ODMR functions, which in practice, can be computed using the efficient FFT techniques. 
Further, uni-variate linear curve-fitting approach is computationally very efficient. 
The super-resolution approach, on the other hand is slightly more computationally complex compared to line-fitting approach as it requires eigenvalue decomposition \cite{moon2000mathematical}.
However, as shall be shown later in Section \ref{secres} on real data, both the proposed approaches are much quicker without any performance loss. 

\section{Results \& Discussion}\label{secres}

Two ODMRs with sampling rate of $100$ kHz about a single resonant frequency, obtained experimentally from the setup described in \cite{setup}, were utilised in studying and evaluating the proposed techniques. 
Both the techniques were implemented using built-in functions of scipy \cite{scipy}. 
We used the time filter width $T_t = 4$ samples, and a threshold of $ 15 \mu T $ and $-15 \mu T$. Further Median Filtering was used on top of it to get the final constructed $B_{src}$ shown in Fig. \ref{map}.  
Fig. \ref{noncur}, \ref{lincur} and \ref{esprit} shows the construction using the existing LF-based, the proposed Phase based linear curve fitting and proposed Phase based ESPRIT approaches respectively. Here, we employed the Trust Region Reflective (TRF) algorithm \cite{lm} for LF based non-linear curve fitting approach. From the figures, one can notice similar  construction using all the three methods. 
However, our implementation of Phase based techniques took $\sim 70 \times$ lesser time than that of LF based method. This huge reduction can be attributed to the efficient computation of FT using FFTs and requiring only a simple curve-fitting unlike the existing LF based approach. Note that as mentioned previously, while the proposed ESPRIT approach requires eigenvalue decomposition, since only 8 samples are considered (time-width = $2T_t$), the increase in complexity compared to the proposed linear curve-fitting approach is marginal and both the proposed approaches yielded results in almost similar time. 

\begin{figure}[ht]
    \centering
    \includegraphics[width=\linewidth]{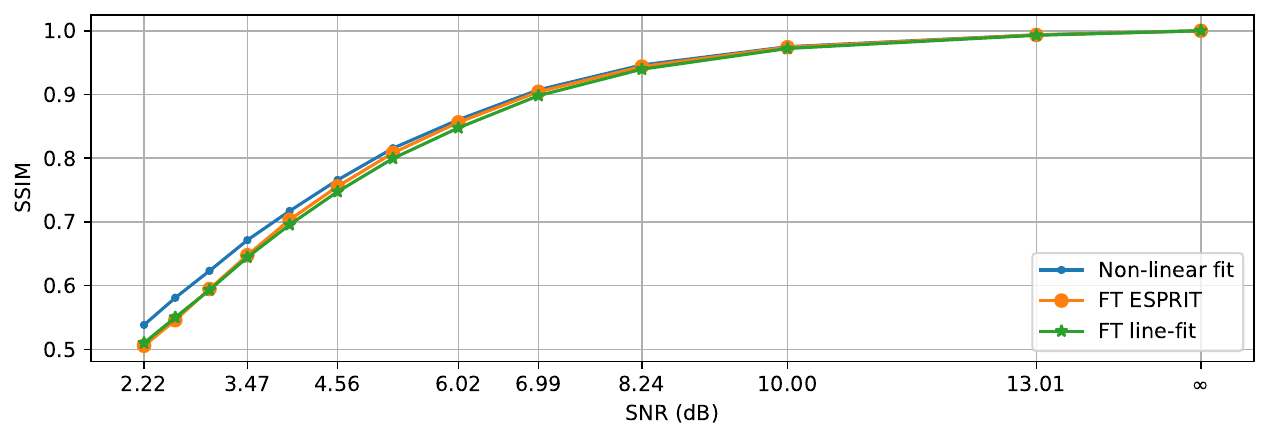}
    \caption{Robustness against the addition of noise: Zero-mean Gaussian noise was added in the ODMR with varying Signal-to-Noise ratio (SNR) and a magnetic field map was produced. This map was compared to the magnetic field map obtained using the original ODMR signal using the respective method.}
    \label{ssim}
\end{figure}

The techniques were further tested for robustness against noise by adding suitable Gaussian noise corresponding to different SNRs. 
The images generated from these noisy ODMRs are compared with the original images constructed through corresponding techniques using structural similarity index measure (SSIM) \cite{sims} which are plotted in Fig. \ref{ssim}. The similarity measure improves as SNR increases with more or less similar characteristics, thus validating an equal robustness of existing LF based and the proposed approaches.

The results hence demonstrates similar performance of the existing and proposed approaches, however with a significant reduction in the computational time, thus making the proposed approach attractive for magnetic field construction from ODMR in NV magnetometry. 

\vspace{1.5mm}
\section{Conclusion}\label{seccon}
NV center-based magnetometry is emerging as a useful technology with different applications. Determining accurate magnetic field maps is one of the important intermittent outputs in the analytics pipeline to carry out the inference relevant to the concerned application. It is interesting to see many of the modules through a signal-processing lens and arrive at useful algorithms/techniques not only for NV magnetometry but also for different quantum sensing systems. This paper proposed a computationally faster technique to construct the magnetic field by bundling some of the techniques regularly used in the signal processing community; an advantage possibly benefiting the development of portable versions of the magnetometers. Driven by this, immense possibility is seen to consider different signal processing techniques not only for magnetic field construction but also for other modules in the pipeline for the application at hand. For instance, one can look into the inverse problem of current reconstruction from the magnetic field maps to charge flow in semiconductor chips, which is useful for different purposes.


\IEEEtriggeratref{7}
\bibliographystyle{IEEEtran}


\end{document}